\documentclass[twocolumn,showpacs,showkeys,preprintnumbers,amsmath,amssymb]{revtex4}
\usepackage{docs}%
\usepackage{bm}%
\usepackage[dvips]{graphicx}
\expandafter\ifx\csname package@font\endcsname\relax\else
 \expandafter\expandafter
 \expandafter
 \usepackage
 \expandafter\expandafter
 \expandafter{\csname package@font\endcsname}%
\fi

\begin{document}

\title{Kinetic Derivation of the Hydrodynamic Equations for Capillary Fluids}

\author{S. De Martino}%
\email{demartino@sa.infn.it} \affiliation{Dipartimento di Fisica
-Universit\`a degli Studi di Salerno, Via S. Allende, Baronissi
(SA), I-84081 Italy; INFM unit\`a di Salerno, 84081 Baronissi
(SA), Italy; INFN, Sezione di Napoli, Gruppo Collegato di
Salerno.}

\author{M. Falanga}%
\email{rosfal@sa.infn.it} \affiliation{Dipartimento di Fisica
-Universit\`a degli Studi di Salerno, Via S. Allende, Baronissi
(SA), I-84081 Italy; INFM unit\`a di Salerno, 84081 Baronissi
(SA), Italy; INFN, Sezione di Napoli, Gruppo Collegato di
Salerno.}

\author{S. I. Tzenov}%
\email{tzenov@sa.infn.it} \affiliation{Dipartimento di Fisica
-Universit\`a degli Studi di Salerno, Via S. Allende, Baronissi
(SA), I-84081 Italy; INFM unit\`a di Salerno, 84081 Baronissi
(SA), Italy; INFN, Sezione di Napoli, Gruppo Collegato di
Salerno.}

\date{May 2004}%



\begin{abstract}
Based on the generalized kinetic equation for the one-particle
distribution function with a small source, the transition from the
kinetic to the hydrodynamic description of many-particle systems
is performed. The basic feature of this new technique to obtain
the hydrodynamic limit is that the latter has been partially
incorporated into the kinetic equation itself. The hydrodynamic
equations for capillary fluids are derived from the characteristic
function for the local moments of the distribution function. The
Fick's law appears as a consequence of the transformation law for
the hydrodynamic quantities under time inversion.
\end{abstract}
\maketitle

\section{Introduction}
\label{sec:intro}

The Van der Waals gradient theory \cite{van} was originally
developed at the end of the 19-th century as an "effective"
physical picture to describe the critical region. This theory
considers a local Helmholtz free-energy composed of two parts. The
first one describes the homogeneous behaviour of the system, while
the second part characterizes the non-homogeneous one. In
addition, the first part is proportional to the fluid density
$\varrho$ while the second one is proportional to the square of
the density gradient. Much effort has been devoted to apply the
Van der Waals gradient theory to various fluids. At present, it
can be considered as the most simple and comprehensive model in
the physics of interfaces and capillarity. The associated
hydrodynamic equations can be written as
\begin{eqnarray}
{\frac {\partial \varrho} {\partial t}} + \nabla \cdot {\left(
\varrho {\bf V} \right)} = 0, \label{Continuity} \\
{\frac {\partial {\bf V}} {\partial t}} + {\left( {\bf V} \cdot
\nabla \right)} {\bf V} = - \nabla {\left[ {\frac {\delta (\varrho
{\cal F})} {\delta \varrho}} \right]}, \label{Korteweg}
\end{eqnarray}
where ${\cal F} {\left( \varrho, \alpha \right)}$ is a function of
the density $\varrho$ and of $\alpha = (1 / 2) {\left| \nabla \rho
\right|}^2$ \cite{anton} and ${\bf V}$ is the current velocity.
This formulation of the Van der Waals theory was originally due to
Korteweg \cite{Korteweg}, who proposed a continuum mechanical
model in which the Cauchy stress tensor apart from the standard
Cauchy-Poisson term contains an additional term defined as

\begin{eqnarray}
{\bf T} = {\left( -p + \alpha \nabla^2 \varrho {\left( {\bf x}; t
\right)} + \beta {\left| \nabla \varrho {\left( {\bf x}; t \right
)} \right|}^2 \right)} {\bf 1} \nonumber \\
+ \delta \nabla \varrho {\left( {\bf x}; t \right)} \otimes \nabla
\varrho {\left( {\bf x}; t \right)} + \gamma {\left( \nabla
\otimes \nabla \right)} \varrho {\left( {\bf x}; t \right)},
\label{Tensor}
\end{eqnarray}
where $\bf 1$ is the unit tensor. As already mentioned by Dunn and
Serrin \cite{du}, the modern terminology concerning the Korteweg
model refers to elastic materials of grade $n$, where the
particular case of $n = 3$ has been well studied in recent years
\cite{slemrod}.

Equations (\ref{Continuity}) and (\ref{Korteweg}) have been linked
recently \cite{an2,de} to a nonlinear Schr\"{o}dinger equation
viewed as a particular set of hydrodynamic equations describing
the so-called nonlinear Madelung fluid \cite{ma,nelson}. This link
between capillarity and the Schr\"{o}dinger equation can shed more
light onto the so-called quantum-like approach to many-particle
systems such as beams in particle accelerators and beam-plasma
systems. The standard procedure in this direction is to
approximate the physical systems characterized by an overall
interaction with a suitable Van der Waals mean field theory.

To avoid misunderstanding, it is worthwhile to note that
Schr\"{o}dinger equation alone does not provide an entire quantum
mechanical picture. It should be necessarily supplemented by a
theory of quantum measurement and consequently by a proper
physical interpretation of wave packets. In the quantum-like
approach, the many-particle systems are described in an effective
way as a whole. Based on the above considerations, it appears
interesting and attractive to explore the possibility of a
rigorous derivation from kinetic theory of the general
hydrodynamic picture thus discussed.

\section{The General Framework}
\label{sec: the general framework}

The starting point of our analysis is the equation for the
microscopic phase space density $N_M {\left( {\bf x}, {\bf p}; t
\right)}$
\begin{equation}
{\frac {\partial N_M} {\partial t}} + {\frac {1} {m}} \nabla \cdot
{\left( {\bf p} N_M \right)} + {\overrightarrow{\partial}}_{\bf p}
\cdot {\left[ {\bf F}_M {\left( {\bf x}, {\bf p}; t \right)} N_M
\right]} = 0, \label{Micphspdens}
\end{equation}
for a system consisting of $N$ particles, which occupies volume
$V$ in the configuration space. Here ${\bf x}$ and ${\bf p}$ are
the coordinates and the canonically conjugate momenta, $m$ is the
particle mass and ${\bf F}_M {\left( {\bf x}, {\bf p}; t \right)}$
is the microscopic force, which apart from the external force
includes a part specifying the type of interaction between
particles. Suppose that at some initial time $t_0$ the microscopic
phase space density is known to be $N_{M0} {\left( {\bf x}, {\bf
p}; t_0 \right)}$. Then, the formal solution of equation
(\ref{Micphspdens}) for arbitrary time $t$ can be written as
\begin{equation}
N_M {\left( {\bf x}, {\bf p}; t \right)} = {\widehat{\cal S}}
{\left( t; t_0 \right)} N_{M0} {\left( {\bf x}, {\bf p}; t_0
\right)}, \label{Formalsol}
\end{equation}
where ${\widehat{\cal S}} {\left( t; t_0 \right)}$ is the
evolution operator, specifying the Hamiltonian flow.

The choice of the initial $N_{M0} {\left( {\bf x}, {\bf p}; t_0
\right)}$ is based on the knowledge of the microscopic
characteristics of the system. Due to the extremely complex
particles' dynamics, full consistent description is not feasible.
Therefore, the detailed information on the microscopic level is
incomplete. If our system is a complex one in the sense that both
the external forces and the collective forces are highly
nonlinear, a dynamic instability of motion is likely to occur on a
characteristic time scale $\tau$. The only information available
to an outside observer by means of a macroscopic measuring device
is a coarse-grained density distribution with a smoothing
function, which takes into account the dynamic instability of
motion. Thus, we assume
\begin{equation}
N_{M0} {\left( {\bf x}; t_0 \right)} = {\widetilde{N}}_M {\left(
{\bf x}; t_0 \right)} = \int {\rm d}^3 {\bf z} G{\left( {\bf x};
t_0 \right| \left. {\bf z} \right)} N_M {\left( {\bf z}; t_0
\right)}, \label{Smoothing}
\end{equation}
where for simplicity the explicit dependence on the momentum
variables ${\bf p}$ has been suppressed. To take into account the
initial preparation of the system, one has to displace the initial
time $t_0$ at $- \infty$ and perform an average over the past
history of the system. Then equation (\ref{Micphspdens}) becomes
\cite{Tzenov}
\begin{eqnarray}
{\frac {\partial N_M} {\partial t}} + {\frac {1} {m}} \nabla \cdot
{\left( {\bf p} N_M \right)} + {\overrightarrow{\partial}}_{\bf p}
\cdot {\left[ {\bf F}_M {\left( {\bf x}, {\bf p}; t \right)} N_M
\right]} \nonumber \\
= {\frac {1} {\tau}} {\left( {\widetilde{N}}_M - N_M \right)}.
\label{Micphspdensms}
\end{eqnarray}
Since the collision time is supposed to be much smaller than the
time $\tau$, the collision integral can be dropped and the kinetic
equation for the one-particle distribution function $f {\left(
{\bf x}, {\bf p}; t \right)}$ can be written as
\begin{equation}
{\frac {\partial f} {\partial t}} + {\frac {{\bf p}} {m}} \cdot
\nabla f + {\bf F} {\left( {\bf x}, {\bf p}; t \right)} \cdot
{\overrightarrow{\partial}}_{\bf p} f = {\frac {1} {\tau}} {\left(
{\widetilde{f}} - f \right)}. \label{Onepartdfsms}
\end{equation}
The right-hand-side of equation (\ref{Onepartdfsms}) is regarded
as a "collision integral", and it can be represented as

\begin{eqnarray}
{\frac {1} {\tau}} {\left( {\widetilde{f}} - f \right)} = \sum
\limits_{l=1}^{\infty} \sum \limits_{{n_1, n_2, \dots, n_k = 0}
\atop{n_1 + n_2 + \dots + n_k = l}}^{l} {\frac {(-1)^l} {n_1! n_2!
\dots n_k!}}\nonumber \\
\times {\frac {\partial^l} {\partial x_1^{n_1} \dots
\partial x_k^{n_k}}} {\left[ {\cal D}_{n_1 n_2 \dots n_k}^{(l)}
{\left( {\bf x}; t \right)} f \right]}, \label{Sdcolintgen}
\end{eqnarray}
where
\begin{eqnarray}
{\cal D}_{n_1 n_2 \dots n_k}^{(l)} {\left( {\bf x}; t \right)} &=&
{\frac {1} {\tau}} \int {\rm d}^3 {\bf z} \Delta z_1^{n_1} \Delta
z_2^{n_2} \dots \Delta z_k^{n_k} G{\left( {\bf z}; t \right|
\left. {\bf x} \right)} \nonumber \\
&=& {\frac {1} {\tau}} {\left \langle \Delta z_1^{n_1} \Delta
z_2^{n_2} \dots \Delta z_k^{n_k} \right \rangle}_{{\bf x},
t}^{(G)}, \label{Moyalcoef}
\end{eqnarray}
with $\Delta {\bf z} = {\bf z} - {\bf x}$. As a first very
interesting step, we consider the diffusion approximation
\begin{equation}
{\frac {1} {\tau}} {\left( {\widetilde{f}} - f \right)} = -
\nabla_k {\left[ A_k {\left( {\bf x}; t \right)} f \right]} +
{\frac {1} {2}} \nabla_k \nabla_l {\left[ B_{kl} {\left( {\bf x};
t \right)} f \right]}, \label{Sdcolintil}
\end{equation}
where
\begin{equation}
A_k ({\bf x}; t) = {\frac {1} {\tau}} {\left \langle \Delta z_k
\right \rangle}_{{\bf x}, t}^{(G)}, \quad B_{kl} ({\bf x}; t) =
{\frac {1} {\tau}} {\left \langle \Delta z_k \Delta z_l \right
\rangle}_{{\bf x}, t}^{(G)}, \label{Sddrdiff}
\end{equation}
and a summation over repeated indices is implied. For Hamiltonian
systems the well-known relation
\begin{equation}
A_k ({\bf x}; t) = {\frac {1} {2}} \nabla_l B_{kl} ({\bf x}; t)
\label{Sddridirel}
\end{equation}
holds, which gives ${\bf A} = 0$ for $B_{kl} = {\rm const}$.

In passing, it is worthwhile to mention that using the principle
of maximum information entropy formulated by Jaynes, it can be
shown \cite{Tzenov} that the smoothing function $G{\left( {\bf z};
t \right| \left. {\bf x} \right)}$ is of the form
\begin{eqnarray}
G {\left( {\left. {\bf z}; t \right|} {\bf x} \right)} = {\frac
{1} {\pi^{3/2} {\sqrt{\left| \det {\widehat{\cal C}} \right|}}}}
\nonumber \\
\times \exp {\left[ - {\left( {\bf z} - {\left \langle {\bf z}
\right \rangle}_{{\bf x}, t}^{(G)} \right)}^T {\widehat{\cal
C}}^{-1} ({\bf x}; t) {\left( {\bf z} - {\left \langle {\bf z}
\right \rangle}_{{\bf x}, t}^{(G)} \right)} \right]}.
\label{Sdsmoothfun}
\end{eqnarray}
The quantities
\begin{equation}
{\left \langle z_k \right \rangle}_{{\bf x}, t}^{(G)}, \qquad
{\left \langle z_k z_l \right \rangle}_{{\bf x}, t}^{(G)},
\label{Sdrestrict}
\end{equation}
are the first and the second moment of ${\bf z}$ at the instant of
time $t + \tau$, provided that ${\bf z}$ measured at the instant
$t$ equals ${\bf x}$ [i.e. ${\bf z} (t) = {\bf x}$]. In addition,
${\widehat{\cal C}} ({\bf x}; t)$ is the covariance matrix defined
as
\begin{equation}
{\cal C}_{kl} ({\bf x}, t) = 2 {\left[ {\left \langle z_k z_l
\right \rangle}_{{\bf x}, t}^{(G)} - {\left \langle z_k \right
\rangle}_{{\bf x}, t}^{(G)} {\left \langle z_l \right
\rangle}_{{\bf x}, t}^{(G)} \right]}. \label{Sdcovarmat}
\end{equation}

The generalized kinetic equation (\ref{Onepartdfsms}) has a form
analogous to the Bhatnagar-Gross-Krook (BGK) equation, widely used
in the kinetic theory of gases \cite{resibois}. There is however,
an important conceptual difference between the two equations. In
the BGK equation the function ${\widetilde{f}}$ should be replaced
by the equilibrium distribution function $f_0$ describing the
global equilibrium and the characteristic time $\tau$ should be
replaced by the corresponding relaxation time. The smoothed
distribution function in equation (\ref{Onepartdfsms})
characterizes a local quasi-equilibrium state within the smallest
unit cell of continuous medium, while $\tau$ is the corresponding
time scale.

\section{The Hydrodynamic Approximation}

Rather than following the standard approach in deriving the
hydrodynamic picture, we introduce the characteristic function
\begin{equation}
{\cal G} {\left( {\bf x}, {\bf w}; t \right)} = \int {\rm d}^3
{\bf p} f {\left( {\bf x}, {\bf p}; t \right)} {\rm e}^{- i {\bf
w} \cdot {\bf p}}, \label{Charfunc}
\end{equation}
instead. It is straightforward to verify that ${\cal G}$ satisfies
the following equation
\begin{equation}
{\frac {\partial {\cal G}} {\partial t}} + {\frac {i} {m}} \nabla
\cdot {\overrightarrow{\partial}}_{\bf w} {\cal G} + i {\bf w}
\cdot {\bf F} {\cal G} = - \nabla \cdot {\left( {\bf A} {\cal G}
\right)} + {\frac {1} {2}} \nabla_n \nabla_s {\left( B_{ns} {\cal
G} \right)}. \label{Charfunceq}
\end{equation}
Note that the local moments ${\left \langle p_1^{n_1} p_2^{n_2}
\dots p_k^{n_k} \right \rangle}$ can be obtained from the
characteristic function according to the relation
\begin{equation}
{\left \langle p_1^{n_1} p_2^{n_2} \dots p_k^{n_k} \right \rangle}
= {\left. i^l {\frac {\partial^l {\cal G}} {\partial w_1^{n_1}
\partial w_2^{n_2} \dots \partial w_k^{n_k}}} \right|}_{{\bf
w}=0}, \label{Localmom}
\end{equation}
where $n_1 + n_2 + \dots + n_k = l$. The well-known hydrodynamic
quantities, such as the mass density $\varrho$, the mean velocity
${\bf V}_{(+)}$ of a fluid element and the hydrodynamic stress
tensor $\Pi_{kl}$ can be defined as
\begin{equation}
\varrho {\left( {\bf x}; t \right)} = m n {\cal G} {\left( {\bf
x}, 0; t \right)} = m n  \int {\rm d}^3 {\bf p} f {\left( {\bf x},
{\bf p}; t \right)}, \label{Hydrorho}
\end{equation}
\begin{equation}
\varrho {\left( {\bf x}; t \right)} {\bf V}_{(+)} {\left( {\bf x};
t \right)} = {\left. i n {\overrightarrow{\partial}}_{\bf w} {\cal
G} \right|}_{{\bf w}=0} = n  \int {\rm d}^3 {\bf p} {\bf p} f
{\left( {\bf x}, {\bf p}; t \right)}, \label{Hydroforvel}
\end{equation}
\begin{equation}
\Pi_{kl} {\left( {\bf x}; t \right)} = - {\frac {n} {m}} {\left.
{\frac {\partial^2 {\cal G}} {\partial w_k \partial w_l}}
\right|}_{{\bf w}=0} = {\frac {n} {m}} \int {\rm d}^3 {\bf p} p_k
p_l f {\left( {\bf x}, {\bf p}; t \right)}, \label{Hydrostrten}
\end{equation}
Here, $n = \lim \limits_{N,V \rightarrow \infty} (N / V)$, implies
the thermodynamic limit . Defining also the deviation from the
mean velocity as
\begin{equation}
m {\bf c}_{(+)} = {\bf p} - m {\bf V}_{(+)}, \label{Devmeanvel}
\end{equation}
and using the evident relation
\begin{equation}
\int {\rm d}^3 {\bf p} {\bf c}_{(+)} {\left( {\bf x}, {\bf p}; t
\right)} f {\left( {\bf x}, {\bf p}; t \right)} = 0, \nonumber
\end{equation}
we can represent the stress tensor $\Pi_{mn}$ according to the
relation
\begin{equation}
\Pi_{mn} = \varrho V_{(+)m} V_{(+)n} + {\cal P}_{mn}.
\label{Internst}
\end{equation}
Here
\begin{equation}
{\cal P}_{kl} {\left( {\bf x}; t \right)} = m n \int {\rm d}^3
{\bf p} c_{(+)k} c_{(+)l} f {\left( {\bf x}, {\bf p}; t \right)},
\label{Internstdef}
\end{equation}
is the internal stress tensor.

Equation (\ref{Charfunceq}) and the one obtained after
differentiating with respect to $w_k$ evaluated at ${\bf w}=0$,
yield the Smoluchowski equation and the equation for the momentum
balance, respectively. These can be written in the form
\begin{equation}
{\frac {\partial \varrho} {\partial t}} + \nabla \cdot {\left[
\varrho {\left( {\bf V}_{(+)} + {\bf A} \right)} \right]} = {\frac
{1} {2}} \nabla_k \nabla_l {\left( B_{kl} \varrho \right)},
\label{Contineq}
\end{equation}
\begin{eqnarray}
{\frac {\partial} {\partial t}} {\left( \varrho V_{(+)k} \right)}
+ \nabla_l {\left( \varrho V_{(+)k} V_{(+)l} \right)} = {\frac
{\varrho} {m}} F_k \nonumber \\
- \nabla \cdot {\left( {\bf A} \varrho V_{(+)k} \right)} -
\nabla_l {\cal P}_{kl} + {\frac {1} {2}} \nabla_l \nabla_n {\left(
B_{ln} \varrho V_{(+)k} \right)}. \label{Momentbal}
\end{eqnarray}
Let us consider the time inversion transformation specified by
\cite{Tzenov,Guerra} $t \rightarrow {\widetilde{t}} = -t$, ${\bf
x} \rightarrow {\widetilde{\bf x}} = {\bf x}$ and ${\bf p}
\rightarrow {\widetilde{\bf p}} = - {\bf p}$. We argue that there
exists a backward velocity ${\bf V}_{(-)} {\left( {\bf x}, t
\right)}$ such that
\begin{equation}
{\widetilde{\bf V}}_{(+)} {\left( {\bf x}, - t \right)} = - {\bf
V}_{(-)} {\left( {\bf x}, t \right)}. \label{Backwvel}
\end{equation}
The transformed Smoluchowski equation (\ref{Contineq}) can be
represented according to
\begin{equation}
{\frac {\partial \varrho} {\partial t}} - \nabla \cdot {\left[
\varrho {\left( - {\bf V}_{(-)} + {\bf A} \right)} \right]} = -
{\frac {1} {2}} \nabla_k \nabla_l {\left( B_{kl} \varrho \right)}.
\label{Contineqb}
\end{equation}
Summing up and subtracting equations (\ref{Contineq}) and
(\ref{Contineqb}), we obtain the continuity equation
\begin{equation}
{\frac {\partial \varrho} {\partial t}} + \nabla \cdot {\left(
\varrho {\bf V} \right)} = 0, \label{Continequa}
\end{equation}
and the Fick's law
\begin{equation}
U_k = - A_k + {\frac {1} {2 \varrho}} \nabla_l {\left( B_{kl}
\varrho \right)}. \label{Ficklaw}
\end{equation}
Here
\begin{equation}
{\bf V} = {\frac {1} {2}} {\left( {\bf V}_{(+)} + {\bf V}_{(-)}
\right)}, \qquad {\bf U} = {\frac {1} {2}} {\left( {\bf V}_{(+)} -
{\bf V}_{(-)} \right)}, \label{Currentosm}
\end{equation}
are the current and the osmotic velocity, respectively. It is
worthwhile to mention that since the mean velocity of a fluid
element is a generic function of time $t$, it can be split into
odd and even part. Note that from equation (\ref{Currentosm}) it
follows that ${\bf V}_{(+)} = {\bf V} + {\bf U}$, where ${\bf V}$
is the odd part, while ${\bf U}$ is the even part.

Equation (\ref{Momentbal}) for the balance of momentum can be
written alternatively as
\begin{eqnarray}
{\frac {\partial V_{(+)k}} {\partial t}} + V_{(-)l} \nabla_l
V_{(+)k} = {\frac {F_k} {m}} \nonumber \\
+ A_l \nabla_l V_{(+)k} - {\frac {1} {\varrho}} \nabla_l {\cal
P}_{kl} + {\frac {B_{ln}} {2}} \nabla_l \nabla_n V_{(+)k}.
\label{Momentbalb}
\end{eqnarray}
After performing a time inversion in equation (\ref{Momentbalb}),
we obtain
\begin{eqnarray}
{\frac {\partial V_{(-)k}} {\partial t}} + V_{(+)l} \nabla_l
V_{(-)k} = {\frac {F_k} {m}} \nonumber \\
- A_l \nabla_l V_{(-)k} - {\frac {1} {\varrho}} \nabla_l
{\widetilde{\cal P}}_{kl} - {\frac {B_{ln}} {2}} \nabla_l \nabla_n
V_{(-)k}, \label{Momentbalbi}
\end{eqnarray}
where ${\widetilde{\cal P}}_{kl}$ denotes the transformed internal
stress tensor after performing the time inversion. Summing up the
last two equations, we arrive at the sought-for equation for the
current velocity
\begin{eqnarray}
{\frac {\partial V_k} {\partial t}} + V_l \nabla_l V_k = {\frac
{F_k} {m}} + A_l \nabla_l U_k \nonumber \\
- {\frac {1} {\varrho}} \nabla_l {\overline{\cal P}}_{lk} + U_l
\nabla_l U_k + {\frac {B_{ln}} {2}} \nabla_l \nabla_n U_k,
\label{Currentvel}
\end{eqnarray}
where
\begin{equation}
{\overline{\cal P}}_{kn} = {\frac {1} {2}} {\left( {\cal P}_{kn} +
{\widetilde{\cal P}}_{kn} \right)}. \label{Inttensor}
\end{equation}
In order to find the explicit form of the internal stress tensor
(\ref{Internstdef}), we observe that the maximum entropy of the
system is realized, provided the small source in the generalized
kinetic equation (\ref{Onepartdfsms}) vanishes. This condition is
equivalent to the condition of detailed balance in the case, where
the collision integral (small source) is approximated by a
Fokker-Planck operator. The condition of detailed balance implies
that the distribution function factorizes in the form
\begin{equation}
f_{eq} {\left( {\bf x}, {\bf p}; t \right)} = {\frac {\varrho
{\left( {\bf x}; t \right)} } {mn}} {\cal F} {\left( {\bf p}; t
\right)}, \label{Factorize}
\end{equation}
where ${\cal F} {\left( {\bf p}; t \right)}$ is a normalizable
function. From the above considerations, it follows directly that
\begin{equation}
{\cal P}_{kl} {\left( {\bf x}; t \right)} = {\frac {3 k_B T} {m}}
\varrho {\left( {\bf x}; t \right)} \delta_{kl},
\label{Internstequi}
\end{equation}
where $k_B$ is the Boltzmann constant and $T$ is the
temperature.\\ In the simplest case, where the external force
vanishes and the diffusion tensor is diagonal and isotropic,
$B_{kl} = \beta \delta_{kl}$, we obtain
\begin{equation}
{\frac {\partial {\bf V}} {\partial t}} + {\left( {\bf V} \cdot
\nabla \right)} {\bf V} = - \nabla {\left( \alpha \ln \varrho -
{\frac {\beta^2} {2}} {\frac {\nabla^2 {\sqrt{\varrho}}}
{\sqrt{\varrho}}} \right)}, \label{Kortewegf}
\end{equation}
where $\alpha = 3 k_B T / m$. Thus, the hydrodynamic equations
describing a free capillary fluid have been recovered. \\ In the
case, where an external force is applied, the Korteweg stress
tensor contains an additional term proportional to the drift
coefficient ${\bf A}$. On the other hand from the principle of
detailed balance, it follows that the drift coefficient is
proportional to the external force. The physical implication of
the latter is that the additional term in the Korteweg stress
tensor can be regarded as a coupling between the external field
and the mean field of purely hydrodynamical origin.

\section{Conclusion}
Since a detailed information about the system on the microscopic
level is incomplete, one possible way to take into account its
initial preparation, i. e. an eventual dynamic instability of
motion that might have set in and/or other large-scale
characteristics, is to introduce a suitable smoothing procedure.
As a result, the kinetic equation providing an unified kinetic,
hydrodynamic and diffusion description contains a small source and
is therefore irreversible. Although the effective collision
integral (small source) can be represented as a Kramers-Moyal
expansion, for the purposes of the present paper it suffices to
consider the right-hand-side of the generalized kinetic equation
as approximated with a properly defined Fokker-Planck operator.
The latter form of the collision term is adopted as a starting
point in the derivation of the hydrodynamic equations for
capillary fluids.\\ The hydrodynamic approximation is further
obtained in a standard manner from the characteristic function for
the local moments of the distribution function. An important
feature of the approach is that the Fick's law emerges naturally
from the transformation properties of the hydrodynamic quantities
under time inversion. The osmotic velocity is uniquely specified
by the first two infinitesimal moments of the smoothing function
and in a sense is a measure of the irreversibility. \\ The main
result of the analysis performed in this paper, the hydrodynamic
equations for free capillary fluids have been derived from kinetic
theory. If an external force is present, the Korteweg stress
tensor has to be modified accordingly. An additional term
proportional to the drift coefficient emerges implying a coupling
between the external field and the mean field of purely
hydrodynamical origin.

\end{document}